# Theoretical Investigation: 2D N-Graphdiyne Nanosheets as Promising Anode Materials for Li/Na Rechargeable Storage Devices

Meysam Makaremi,[1] Bohayra Mortazavi,[2] Timon Rabczuk,[2] Geoffrey A. Ozin,[3] and Chandra Veer Singh*,[1,4]

[1]Department of Materials Science and Engineering, University of Toronto, 184 College Street, Suite 140, Toronto, ON M5S 3E4, Canada.
[2]Institute of Structural Mechanics, Bauhaus-Universität Weimar, Marienstr. 15, D-99423 Weimar, Germany.
[3]Department of Chemistry, Solar Fuels Research Cluster, University of Toronto, 80 St. George Street, Toronto, ON M5S 3H6, Canada.
[4]Department of Mechanical and Industrial Engineering, University of Toronto, 5 King's College Road, Toronto M5S 3G8, Canada.

**Abstract**

N-graphdiyne monolayers, a set of carbon-nitride nanosheets, have been synthesized recently through the polymerization of triazine- and pyrazine-based monomers. Since the two-dimensional nanostructures are mainly composed of light-weight nonmetallic elements including carbon and nitrogen, they might be able to provide high storage capacities for rechargeable cells. In this study, we used extensive first principle calculations such as electronic density of states, band structure, adsorption energy, open-circuit voltage, nudged-elastic band and charge analyses to investigate the application of the newly fabricated N-graphdiyne monolayers as the anode material for Li/Na/Mg ion batteries. Our calculations suggest that while Mg foreign atoms poorly interact with monolayers, Li and Na adatoms illustrate outstanding anodic characteristics for rechargeable storage cells. Electronic density of states calculations indicate that the insertion of Li/Na into the novel N-graphdiyne materials enhances the electrical conductivity of nanosheets. Adsorption energy and open-circuit voltage calculations predict that the nanosheets can provide a high storage capacity spectrum of 623 – 2180 mAh/g which is higher than that for most recently discovered 2D materials (*e.g.* phosphorene, borophane, and germanene involve Li binding capacities of 433, 504, and 369 mAh/g, respectively) and it is also significantly greater than the capacity of commercial anode materials (*e.g.* graphite contains a capacity of 372 mAh/g). This study provides valuable insights about the electronic characteristics of newly fabricated N-graphdiyne nanomaterials, rendering them as promising candidates to be used in the growing industry of rechargeable storage devices.

KEYWORDS

N-graphdiyne, Li/Na ion storage, anode electrode, electronic structure, storage capacity, open circuit voltage.

## 1. Introduction

Rechargeable Lithium-ion battery (LIB) has been one of the most widely used electronic energy storage devices in recent years. LIB has been the main research focus in many governmental multidisciplinary labs and international scientific institutes, and it is a key component for home, portable and aerospace electronics and outperforms other types of storage cells; such as, nickel-cadmium and nickel metal hydride batteries by supplying higher charge storage capacity, lower maintenance expense and longer lifetime.[1–4] In LIB, Li ions move from/to the anode to/from the cathode during the discharge/charge process, and this way the characteristics of the electrode's active materials play critical role on the overall performance of the rechargeable batteries.[5]

In the past years extensive efforts have been devoted to discover satisfactory an anode material for rechargeable ion batteries. Graphite is one of the most common anodes for LIB which includes robust stability, electrical conductivity, and low cost; whereas, it suffers from the low charge capacity of 372 mAh/g.[6–8] Lithium titanium oxide (LTO) is another commercialized anode material which consists of outstanding life cycle and thermal stability; though it suffered from a low capacity (175 mAh/g), and a high cost.[9] Silicon with a very high theoretical storage capacity (up 4200 mAh/g) was suggested as an alternative for graphite, although it undergoes a massive volume expansion/contraction (up to 300%) due to lithiation/delithiation resulting in the structural degradation.[10–12] On the other hand, among different anode candidates, 2D materials have been offered as reliable alternatives with a large surface-to-volume ratio leading to outstanding magneto-electronic, mechanical, and thermal aspects.[13–17] Studies performed in past few years have highlighted the applicability of 2D materials as anodes materials for rechargeable batteries, which involve remarkable storage capacities[18–21] and fast diffusion rates.[22,23]

In last decade several 2D structures have been predicted, fabricated and investigated. 2D nanosheets might be fabricated by top-down and bottom-up approaches,[24] and they might consist of a single element; such as graphene, silicene, and germanene, or they may contain several elements such as metal nitrides/carbides (MXenes)[25,26], transition metal dichalcogenides (TMDs)[27–30], and 2D metal-organic frameworks (MOFs).[31,32] Lately, a new family of 2D crystals called carbon nitride (CN) nanosheets; such as $C_2N$[33], $C_3N$[34], $C_3N_4$[35], and nitrogen-graphdiyne[36–38] which are mainly composed of carbon and nitrogen atoms have been experimentally fabricated,[39] and more CN sheets with varying stoichiometry and structures are theoretically studied.[40–42] CN nanosheets enclose uniform atomic networks containing C and N elements with comparable atomic radii composed of four and five valence electrons, respectively, which can form robust covalent structures with promising opto-electronic properties for green energy applications. [43–45] Recently, Kan *et al.*[46] synthesized three new members of the 2D CN family which contain conjugated N-graphdiyne structures by polymerization of triazine- or pyrazine-based monomers with ethynyl group via Glaser coupling reactions at the gas/liquid and liquid/liquid interfaces. The atomic structures of 2D N-graphdiyne monolayers were identified by TEM, SEM, AFM, XPS, and Raman spectra approaches offering that the newly fabricated nanosheets resemble the graphyne configurations; though, two or three carbon atoms of connecting hexagonal rings are substituted with nitrogen atoms.

The presence of light-weight elements; such as carbon and nitrogen, in the newly synthesized nanostructures, and also the outstanding optoelectronic properties of previously discovered graphyne/graphdiyne monolayers as anode materials[47–49] propose this novel class of 2D materials as promising candidates to design advanced batteries anode electrodes with high storage capacities. In this work we calculated various anodic aspects of N-graphdiyne crystals due to the intercalation of Li, Na and Mg foreign atoms; such as anode storage capacity, electronic charge density, electronic density of states, open-circuit voltage, and ion diffusion characteristics through carrying out comprehensive DFT simulations. The adatoms were gradually inserted and optimum charge capacity of CN monolayers including the maximal charge transfer were obtained, and *ab-initio* molecular dynamics (AIMD) simulations were carried out to analyze the thermal stability of the novel CN crystals. We also studied the transfer of metallic adatoms over the nanosheets by using nudged-elastic band calculations.

## 2. Computational Details

The spin-polarized density functional theory (DFT) method with a Grimme dispersion correction technique, DFT-D2[50] as implemented in the Vienna Ab-initio Simulation Package (VASP)[51] was used via generalized gradient approximation (GGA) with Perdew–Burke–Ernzerhof (PBE)[52] and Heyd-Scuseria-Ernzerhof (HSE06)[53] functionals. The projector augmented-wave (PAW) technique[54] was applied to define the interaction between valence and core electrons. The kinetic energy cutoff, the electronic self-consistency criterion, and ionic relaxation convergence threshold were considered to be 500 eV, 1×10-6 eV and 1×10-3 eV/Å, respectively. To visualize equilibrated structures and charge densities the VESTA[55] package was leveraged. A Monkhorst-Pack[56] mesh size of 15×15×1 and the tetrahedron scheme with Blöchl corrections were used for the integration of the Brillouin zone. The electronic charge gain and loss were estimated by the Bader summation approach[57]. The system charge difference due to adsorption of an adatom can be described as,

$$\Delta\rho = \rho_{Total} - \rho_{CN} - \rho_M , \qquad (1)$$

where, $\rho_{Total}$, $\rho_{CN}$, and $\rho_M$ represent the charge densities of sheet+adatom, sheet, and the adatom, respectively. The adsorption energy due to the binding of adatoms is defined by,

$$E_{Bind} = \frac{(E_{Total} - E_{CN} - n \times E_M)}{n} , \qquad (2)$$

where $E_{Total}$, $E_{CN}$, $E_M$ involve the total interaction energy, the energy of the monolayer, and the energy of the individual adatom in the gas phase; respectively, and n is the number of inserted adatoms. A negative amount for $E_{Bind}$ shows the binding between the nanosheets and adatoms.

Figure 1 illustrates the unitcell configurations for N-graphdiyne monolayers which include 24 (18 C and 6 N), 28 (24 C and 4 N), and 42 (36 C and 6 N) atoms; respectively, and we call them $C_{18}N_6$, $C_{24}N_4$, and $C_{36}N_6$ accordingly, consistent with a previous study[58,59]. The first and last crystals contain the hexagonal unitcell with the lattice constants of 16.04 Å and 18.66 Å, respectively, while the second monolayer composed of a rectangular lattice with a = 15.97 Å and b = 9.67 Å. Structural details for all three N-graphdiyne crystals are listed in Figures S1 to S6. The A vacuum distance of 20 Å was considered in the z direction (normal to the nanosheet) to prevent image-image interactions along the vacuum space. The periodic boundary conditions were applied for all directions.

**3. Results and Discussion**

Atomic configurations including electronic localization function (ELF)[60] contours of the recently synthesized N-graphdiyne monolayers are illustrated in Figure 1. It is worth pointing out that $C_{24}N_4$, and $C_{36}N_6$ structures involve a similar N-to-C ratio of 1/6; whereas, $C_{18}N_6$ embeds the ratio of 1/3. According to Becke and Edgecombe's definition[61], the red (ELF = 1), green (ELF = 0.5), and blue (ELF = 0.0) contours represent electron localization, electron delocalization and low electron density regions; respectively. As illustrated in Figure 1, the ELF values for each C-C and N-C bond of N-graphdiyne structures is larger than 0.8 which represents the covalent nature of these bonds (a value larger than 0.65 defines covalent bonding). It should also be noted that there is charge localization on each N atom due to its larger number of valence electrons compared to that of carbon. Furthermore, the center of all rings including hexagonal, rhombus, and triangular profiles are completely free from the electron charge which might make them host candidates for the adsorption of interlayer species. With respect to the cohesive energy, $C_{24}N_4$ was found to show higher energetical stability than $C_{36}N_6$. Among the considered monolayers, $C_{18}N_6$ exhibit slightly lower energetical stability than $C_{36}N_6$.[58]

To study anode property of the newly fabricated carbon nitride monolayers, the interaction of the nanosheets with three different foreign atoms including Li, Na, and Mg was investigated. At the beginning, by using binding energy calculations the adsorption behaviors of nanosheets due to the single adatom insertion were surveyed and the most stable adsorption sites containing lowest adsorption energies were uncovered (see Figure 2). The adsorption energies are listed in Table 1.

The first atomic adsorption mostly happens inside the monolayer at the corner of the hexagonal (for $C_{18}N_6$ and $C_{36}N_6$) or the rhombus (for $C_{24}N_4$) ring between the N and C atoms of the monolayer. The only exception takes place for the insertion of Na into the $C_{36}N_6$ nanosheet through which Na binds on the top of the triangular ring. The strongest interactions happen between the Li adatom and N-graphdiyne monolayers (the interaction with the $C_{24}N_4$ nanosheet containing an energy of -2.86 eV is the most vigorous); on the other hand, the intercalation of Mg into the monolayers leads to weakest interactions with respect to the Li and Na adsorption. The difference in the binding intensity of the adatoms can be explained by the difference in the ratio of atomic energy density to the atomic mass. In another word, Li and Mg which respectively contain the largest and the lowest density-to-mass present the strongest and weakest reactivity, accordingly.

The open-circuit voltage (V)[62] of an anode electrode is a key parameter showing the performance of the electrode and can be determined by,

$$V \approx \frac{\left((x_2-x_1)E_{M_B} - (E_{M_{x_2}CN} - E_{M_{x_1}CN})\right)}{(x_2-x_1)e}, \quad (3)$$

here, $x_1$, $x_2$ involve the initial and final adatom coverage ratios. $E_{M_{x_1}CN}$, $E_{M_{x_2}CN}$, and $E_{M_B}$ are the interaction energy for the entire system with coverage ratio of $x_1$ and $x_2$, and the metallic bulk energy; respectively, and e is the electron charge. It should be noted that in our calculations, the reference energies for the Li and Na adatoms were considered to be those of their bulk lattices and not the atomic energies in the vacuum. This equation therefore means that whenever intercalating atoms prefer clustering rather than the binding to the surface, the adsorption energy would become negative, which is the main criteria in determining the maximal energy storage capacity. Table 2 illustrates the anode potential due to the insertion of single adatoms. While the intercalation of Li and Na to different monolayers leads to negative potentials, the Mg insertion results in positive voltages which means this adatom cannot be a proper ion to be employed as electron donor for the N-graphdiyne anode. The charge density contours of Figure 2 also confirm the limited charge donation of Mg to N-graphdiyne monolayers. Therefore, in the following, we investigate the application of Li and Na adatoms for the metallic ion battery usage.

The Li/Na adatoms were gradually and randomly inserted to each monolayer at predefined binding points through the uniform distribution until the monolayer reached to its final capacity point after which there was no more charge transfer to the surface. It is worth nothing that the intercalation and deintercalation of Li/Na ions in cathode and anode electrodes can leads to structural deflections. To ensure about the material performance as an electrode for the rechargeable battery application, induced deformations upon the ions transfer must be limited and may not cause structural damage or cracking. We conducted the AIMD simulations including the largest number of adatoms adsorbed over the both sides (maximum charge capacity) for 20 ps at different temperatures to monitor the structural stability of the N-graphdiyne nanosheets at elevated conditions. Notably, the AIMD results shown in Figure 3 clearly confirm that the all original covalent bonds for N-graphdiyne nanosheets containing the maximal adatom coverage were kept completely intact, even at the elevated temperature of 1000 K. These calculations predict that N-graphdiyne monolayers present remarkable structural and thermal stability as anode materials for Li/Na ion batteries. It should be noted that the pristine monolayers have been reported to be stable up to 2500 K by a previous study[58].

The battery storage capacity of LIB can be calculated via Faraday equation;

$$q = 1000 \, F \, z \frac{n_{max}}{M_{CN}}, \qquad (4)$$

here, F, z, $n_{max}$, and $M_{CN}$ are the Faraday coefficient, adatom valence number, optimal adatom coverage, and the total mass of the nanosheet. The Li/Na storage capacities of different N-graphdiyne nanosheets are presented in Table 3. The intercalation of the Li and Na foreign atoms leads to outstanding capacity spectra of 1,660 - 2,180 mAh/g, and 623 - 934 mAh/g; respectively. The remarkable storage capacities of the newly fabricated carbon nitride crystals can be explained by the porous atomic lattices, presence of constituent light-weight atoms such as N and C in the structures, and also the specific atomic arrangements which enable the monolayers to adsorb interlayer species more effectively.

Plus, among the nanosheet $C_{24}N_4$ illustrates the highest capacities for both Li and Na adsorption including 2,180 mAh/g and 934 mAh/g; respectively, while the atomic adsorption on $C_{36}N_6$ results in the lowest capacities of 1,660 mAh/g and 623 mAh/g; accordingly.

It is worth noting that N-graphdiyne monolayers outperform other anode candidates for LIB including commercial anode materials; such as graphite and $TiO_2$ with capacities of 372 mAh/g and 200 mAh/g; respectively,[63–65] and most of previously synthesized and theorized 2D materials; such as phosphorene, borophane, silicene, germanene, stanene and $Cu_3N$ monolayers showing Li binding capacities of 433, 504, 954, 369, 226, 1008 mAh/g, accordingly.[66–69] Carbon ene-yne (CEY) graphyne[70] entirely composed of carbon atoms is the only exception among LIB anode candidates which shows a higher capacity than N-graphdiyne 2D crystals, including the capacity of 2,680/1,788 mAh/g for Li/Na adsorption. Although, it should be noted that N-graphdiynes display a comprehensive thermal stability up to 2500 K, whereas CEY has been found to be stable only up to 1500 K. [70]

High storage capacity alone would not be advantageous if a rechargeable battery cannot properly convey the stored electrons. Limited electrical conductivity due to internal resistance is a bottleneck causing the ohmic loss and heat generation during charge/discharge cycles. Appropriate electrode materials for the LIB application may require to present metallic band structure during the battery operation.[71] Figure 4 shows the electronic density of states (DOS) of different pristine N-graphdiyne nanosheets, and the ones embedding Li/Na adatoms. Although, at the beginning the pristine $C_{18}N_6$, $C_{24}N_4$, and $C_{36}N_6$ crystals exhibit semiconducting behavior (involving the energy gaps of 2.20, 0.50 and 1.10 eV, respectively), they all develop the metallic band gap after the insertion of one Li/Na foreign atom.

The binding energy and open circuit voltage of the nanosheets as functions of the insertion coverage ratio, and the storage capacity are illustrated in Figure 5. Consistent with literature studies for other 2D crystals [65,70,72], newly fabricated N-graphdiynes show more negative adsorption energies and therefore stronger interactions due to the binding of Li adatoms compared to the Na counterparts. This can be a consequence of the extremely small atomic radius, and high reactivity of Li, which also cause it to be adsorbed by the surface at closer distances. It is worth mentioning that although the initial potentials for Na adsorption are similar or higher than the ones for the Li binding, the Li potential curves decrease more slowly and cause much higher capacities due to the Li binding (more than twice of Na atomic adsorptions). The maximal numbers of Li/Na adsorbed adatoms on $C_{18}N_6$, $C_{24}N_4$, and $C_{36}N_6$ surfaces include 20/8, 28/12, and 32/12, respectively. Worthy to note that in real applications, after first cycles due to the formation of solid/electrolyte interface (SEI) layer the voltage curve may drop. Such a cycling drop in the voltage profile depends on various factors, like the electrolyte chemistry.

As can be seen in Figure 6, nudged elastic band (NEB) calculations were also performed to study atomic diffusion over the $C_{24}N_4$ nanosheet as a representative for N-graphdiynes. There are two possible atomic pathways (Path 1 and Path 2) through which foreign atoms can move. The energy barriers for the Li/Na atomic transfer over Path 1 and Path 2 are 0.80/0.79 eV and 0.95/0.97 eV; respectively, which are a little higher than the of commercial anode materials; such as graphite[73] and TiO$_2$[74] with diffusion barriers of 0.4 eV and 0.65 eV, respectively. It is worth noting that the diffusion energy barrier on phosphorene[75] is 0.13–0.76 eV, and on GeS[76], graphene[77], TiC$_2$[78] and Ti$_3$C$_2$[79] is 0.19 eV, ~0.37 eV, 0.11 eV and ~0.70 eV; respectively. The predicted energy barriers for the Li and Na diffusion over the N-graphdiyne monolayer are located in the upper bound of the range for other 2D materials, which highlights that the N-graphdiyne nanosheets may show limitations to achieve fast charging rates in comparison with other 2D counterparts. It should be noted that in most of previously discovered 2D structures; such as hydrogen boride[65], borophane[67], silicene/germanene/stanene[80], the Li diffusion leads to a larger barrier than the Na movement; however, the N-graphdiyne crystal involves similar energy barriers for both Li and Na moves, consistent with the CEY[70] nanosheet.

## 4. Conclusions

Rechargeable Li ion battery has been a key component of recent electronic devices including communication, aerospace, home and portable electronics. Since current traditional anode materials used in rechargeable LIB suffer from low charge capacity, novel two-dimensional materials with a large surface-to-volume ratio has been suggested to resolve this issue. In this regard, we employed first-principle spin-polarized DFT-D2 calculations to study a set of recently synthesized 2D carbon nitrides called N-graphdiyne monolayers including three stochiometric formulae of $C_{18}N_6$, $C_{24}N_4$, and $C_{36}N_6$.

Various calculations; such as, binding energy, open-circuit potential, nudged-elastic band, density of states, band structure, and Bader charge analyses were utilized to probe the anode properties of the novel 2D structures for Li/Na/Mg ion battery cells. The adsorption of Li/Na adatoms improve the electrical conductivity of N-graphdiyne monolayers, which plays important role for the LIB application. Our calculations illustrate that the configuration of N and C atoms in N-graphdiyne structures can result in a high storage capacity spectrum of 1660 - 2180 mAh/g for the Li ion battery, and a pretty high capacity range of 623 - 934 mAh/g for the Na ion battery.

Comparing with commonly used anode materials; such as graphite with the storage capacity of 372 mAh/g, this study highly recommends the application of the newly fabricated 2D N-graphdiyne crystals as outstanding electrodes for the next generation of Li/Na ion batteries. Moreover, this study will hopefully open up new routes to investigate the application of this novel class of 2D materials as electrodes in rechargeable batteries. In future studies one can probe the role of adatoms concentration on the diffusion rate over the N-graphdiyne monolayers, and also study the effects of layer interactions on the anodic properties of N-graphdiyne multilayers.

AUTHOR INFORMATION

**Corresponding Author**

* chandraveer.singh@utoronto.ca

ACKNOWLEDGMENTS

MM and CVS gratefully acknowledge their financial support in parts by Natural Sciences and Engineering Council of Canada (NSERC), University of Toronto, Connaught Global Challenge Award, and Hart Professorship. The computations were carried out through Compute Canada facilities, particularly SciNet and Calcul-Quebec. SciNet is funded by the Canada Foundation for Innovation, NSERC, the Government of Ontario, Fed Dev Ontario, and the University of Toronto, and we gratefully acknowledge the continued support of these supercomputing facilities. BM and TR greatly acknowledge the financial support by European Research Council for COMBAT project (Grant number 615132).

Table 1: Binding energy of single adatom insertion into N-graphdiyne monolayers.

|  | Li | Na | Mg |
|---|---|---|---|
| $C_{18}N_6$ | -2.61 eV | -2.09 eV | -0.83 eV |
| $C_{24}N_4$ | -2.86 eV | -2.45 eV | -0.97 eV |
| $C_{36}N_6$ | -2.77 eV | -2.55 eV | -0.72 eV |

Table 2: Open-circuit voltage of single adatom insertion into N-graphdiyne monolayers.

|  | Li | Na | Mg |
|---|---|---|---|
| $C_{18}N_6$ | -0.90 V | -0.89 V | 0.96 V |
| $C_{24}N_4$ | -1.15 V | -1.25 V | 0.82 V |
| $C_{36}N_6$ | -1.06 V | -1.34 V | 1.07 V |

Table 3: Li and Na ion Battery storage capacity for N-graphdiyne monolayers.

|  | Li | Na |
|---|---|---|
| $C_{18}N_6$ | 1785 mAh/g | 714 mAh/g |
| $C_{24}N_4$ | 2180 mAh/g | 934 mAh/g |
| $C_{36}N_6$ | 1660 mAh/g | 623 mAh/g |

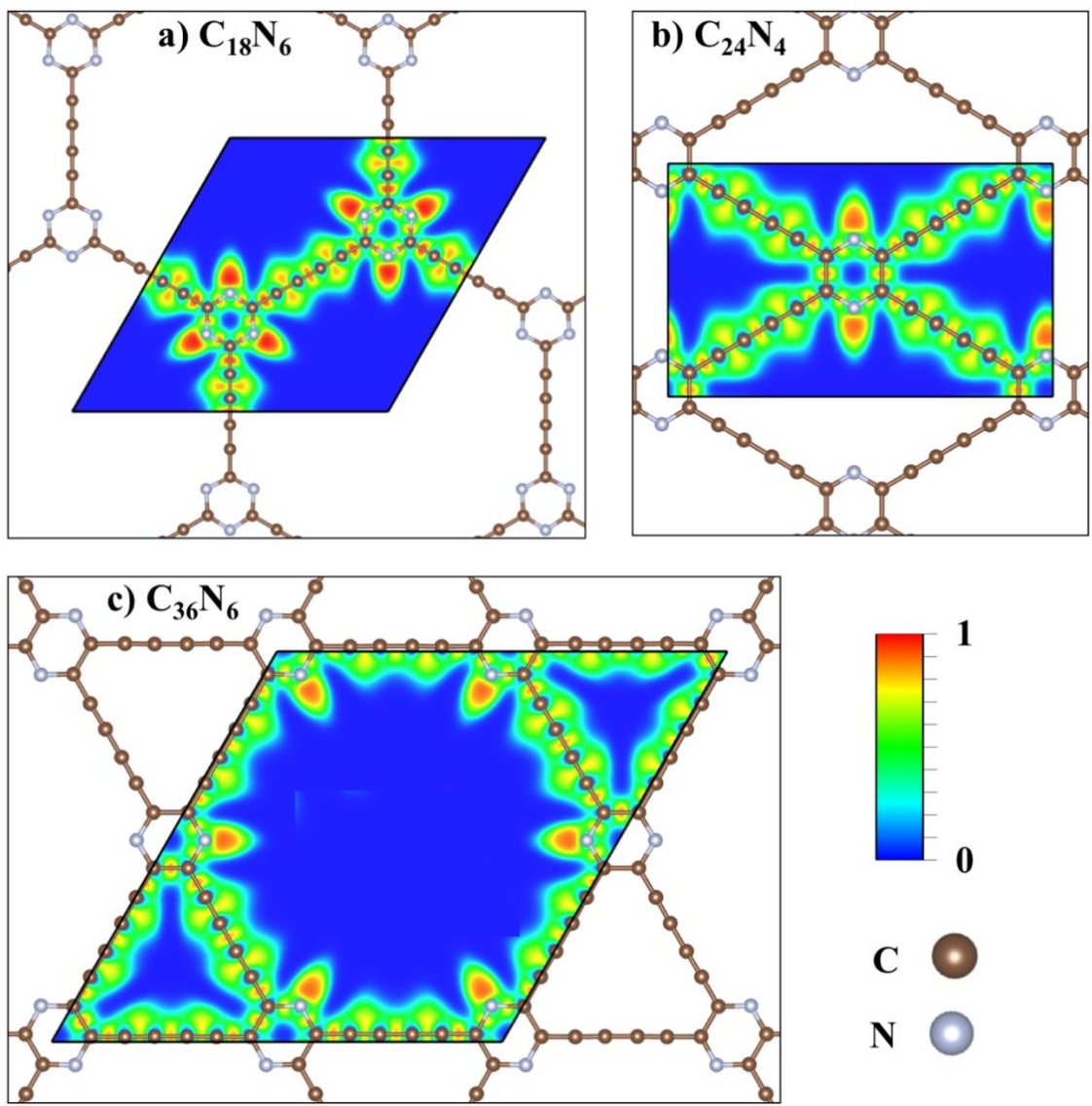

Figure 1. Unitcells of N-graphdiyne nanosheets including a) $C_{18}N_6$ including the hexagonal unitcell with the lattice constants of 16.04 Å, b) $C_{24}N_4$ involving the hexagonal unitcell with the lattice constants of 18.66 Å and c) $C_{36}N_6$ composed of a rectangular unitcell with a = 15.97 Å and b = 9.67 Å.

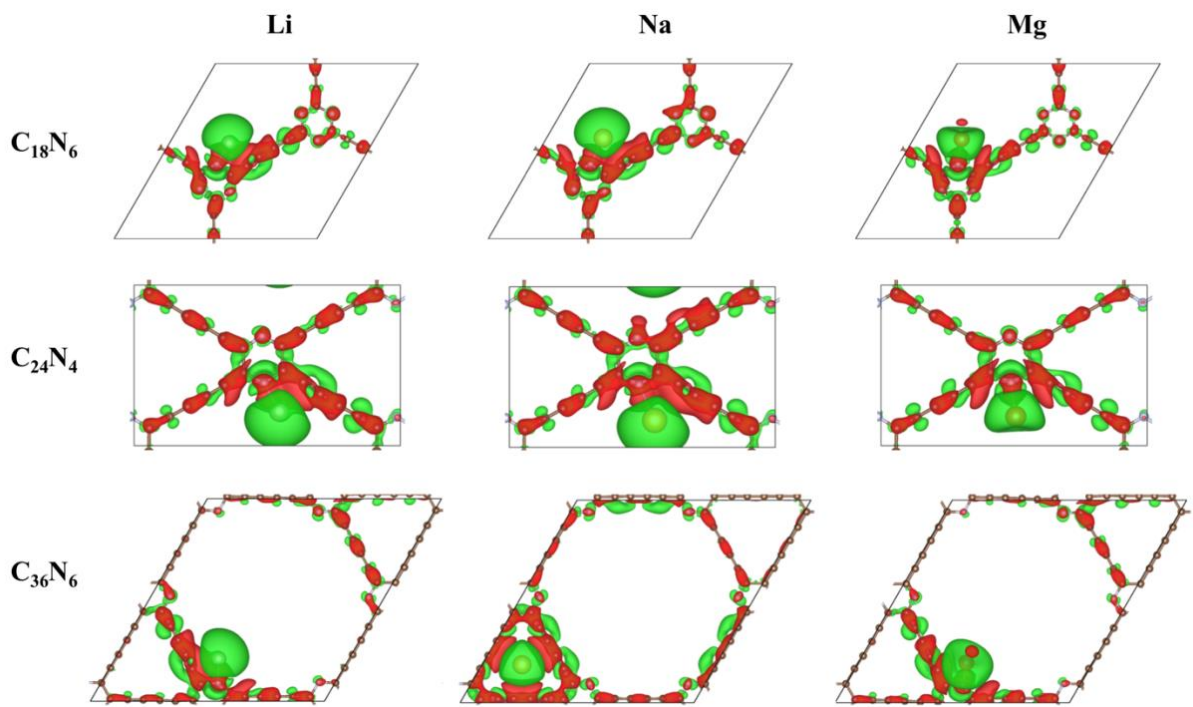

Figure 2. Electronic charge transfers due to the adsorption of Li, Na, and Mg adatoms to different N-graphdiyne monolayers ($C_{18}N_6$, $C_{24}N_4$, and $C_{36}N_6$). Color coding of contours involves green for charge sufficient and red for the charge deficient regions; respectively, (isosurface value is 0.001 e/Å$^3$).

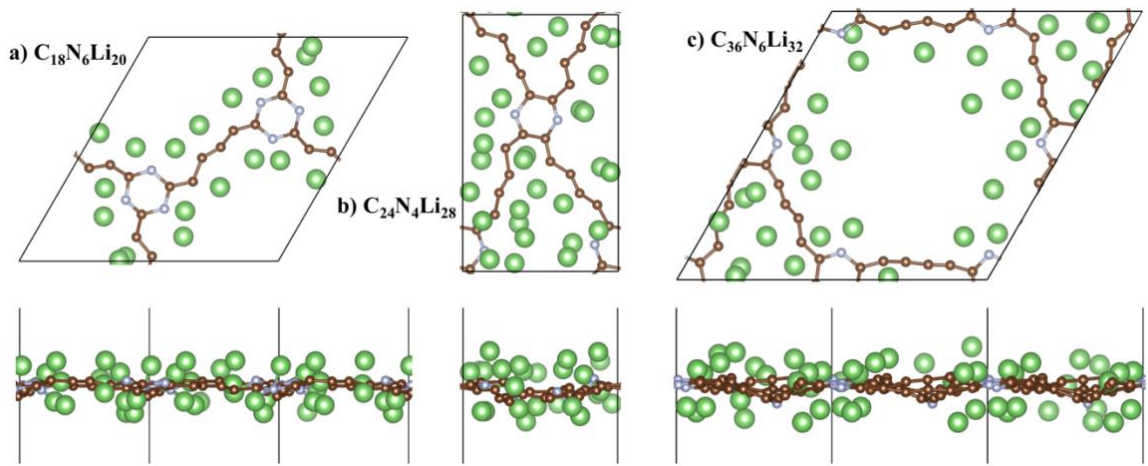

Figure 3. Thermally stable structures of different Li intercalated N-graphdiyne monolayers ($C_{18}N_6$, $C_{24}N_4$, and $C_{36}N_6$) at 1000 K. Brown, gray, and green spheres show C, N, and Li atoms; respectively.

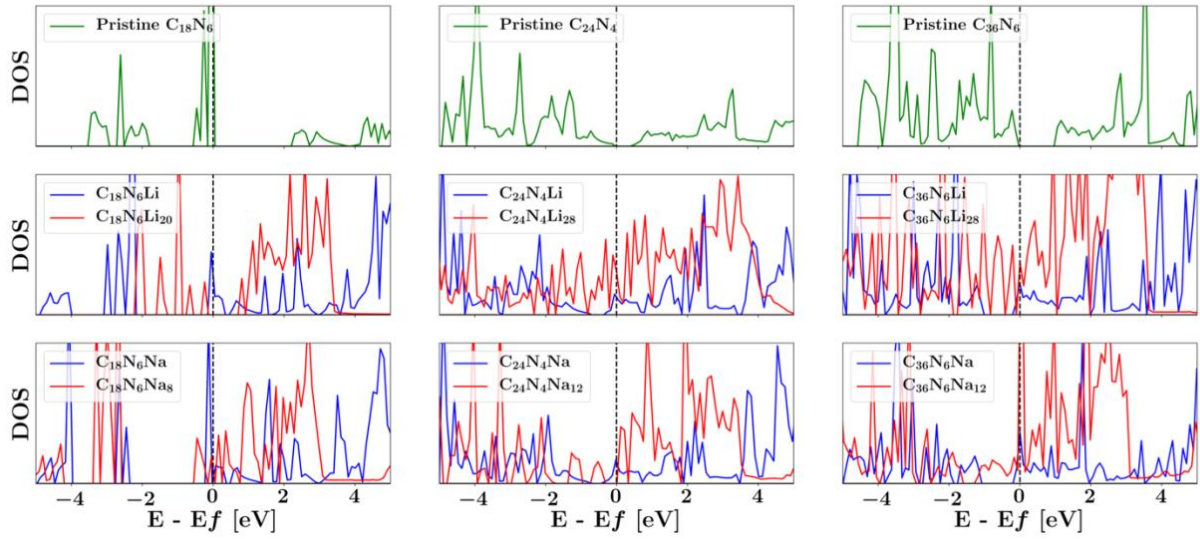

Figure 4. Electronic density of states (DOS) for different pristine N-graphdiyne nanosheets, and the monolayers interacting with various coverages of foreign Li and Na adatoms. The black dashed line represents the Fermi-level.

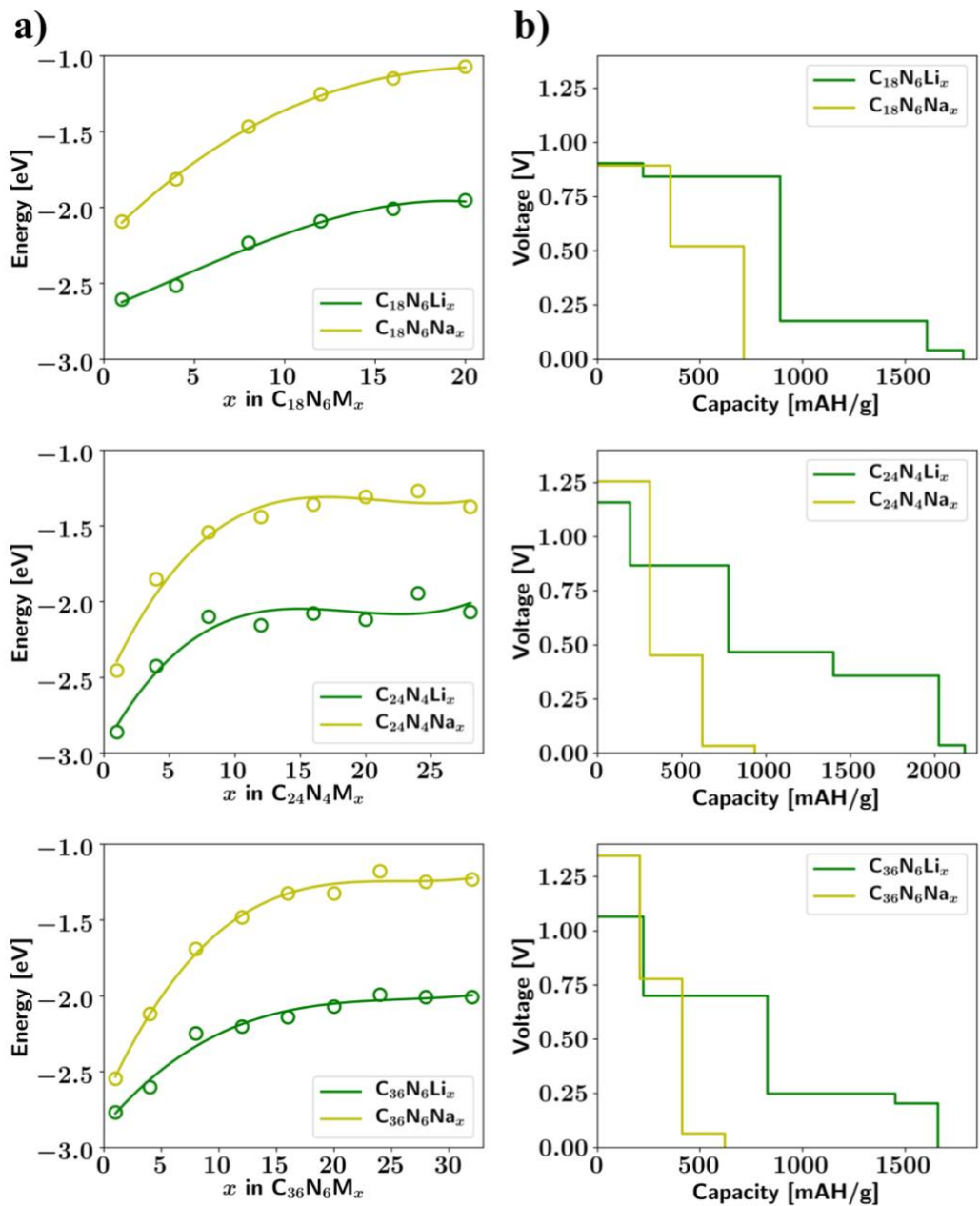

Figure 5. Energetic properties of N-graphdiyne monolayers ($C_{18}N_6$, $C_{24}N_4$, and $C_{36}N_6$) due to the insertion of Li and Na adatoms; a) Adsorption energy with respect to the coverage ratio of x. b) Average open-circuit voltage of the nanosheet as a function of the storage capacity.

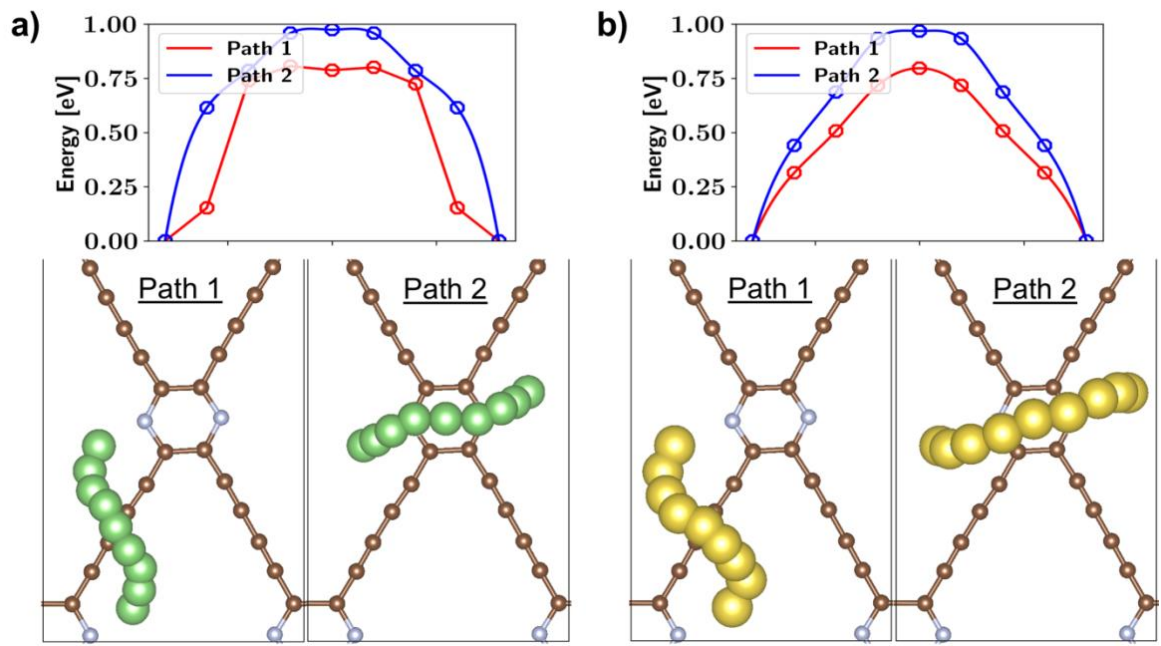

Figure 6. Diffusion of single adatoms over $C_{24}N_4$ monolayer through different pathways (Path 1 and Path 2). a) Li interaction and b) Na intercalation. Brown, gray, green, and yellow spheres represent C, N, Li, and Na atoms; respectively.

For Table of Contents Only

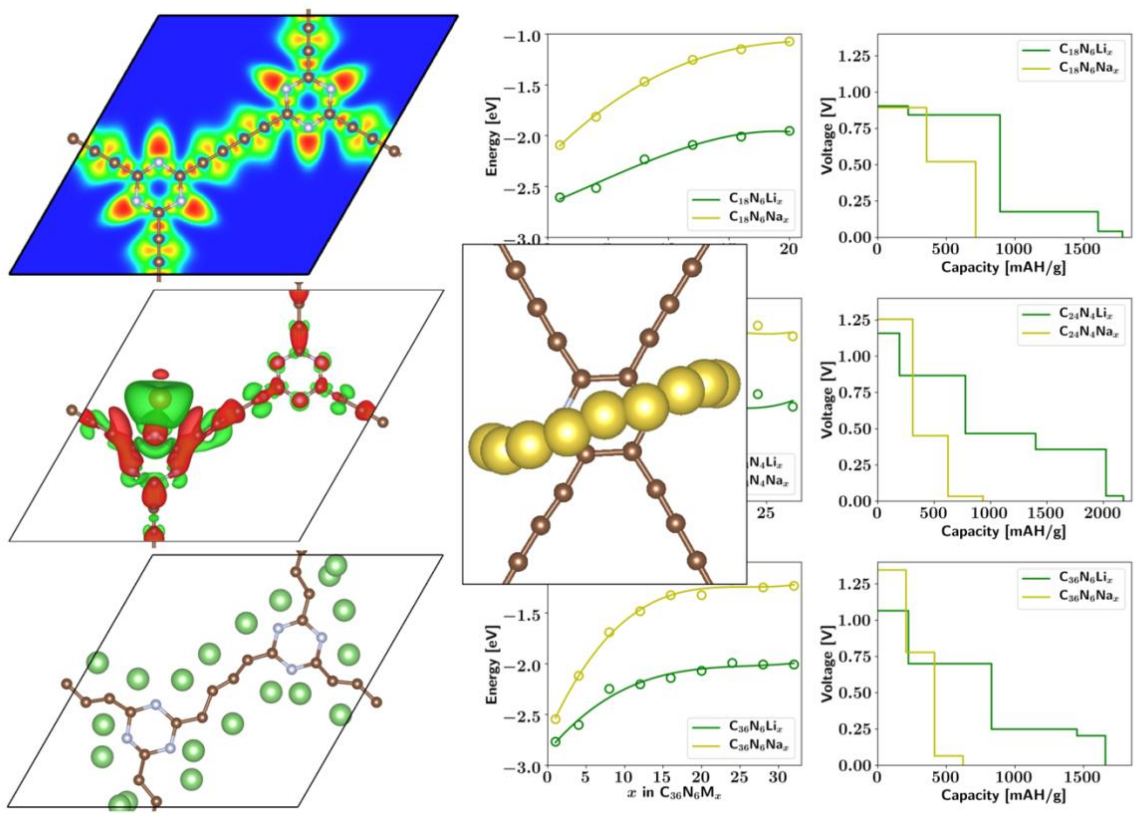